# Understanding resonance graphs using Easy Java Simulations (EJS) and why we use EJS

Loo Kang WEE[1], Tat Leong LEE[1], Charles CHEW[2], Darren WONG[3] and Samuel TAN[1]

[1]Ministry of Education, Educational Technology Division, Singapore
[2]Ministry of Education, Academy of Singapore Teachers, Singapore
[3]Ministry of Education, Curriculum Planning & Development Division, Singapore

lawrence_wee@moe.gov.sg, lee_tat_leong@moe.gov.sg, charles_chew@moe.gov.sg, darren_wong@moe.gov.sg, samuel_tan@moe.gov.sg

Abstract:

This paper reports a computer model- simulation created using Easy Java Simulation (EJS) for learners to visualize how the steady-state amplitude of a driven oscillating system varies with the frequency of the periodic driving force. The simulation shows (N=100) identical spring-mass systems being subjected to (1) periodic driving force of equal amplitude but different driving frequencies and (2) different amount of damping. The simulation aims to create a visually intuitive way of understanding how the series of amplitude versus driving frequency graphs are obtained by showing how the displacement of the system changes over time as it transits from the transient to the steady state.

A suggested "how to use" the model is added to help educators and students in their teaching and learning, where we explained the theoretical steady state equation, time conditions when the model starts allowing data recording of maximum amplitudes to closely match the theoretical equation and steps to collect different runs of degree of damping.

We also discuss two design features in our computer model: A) displaying the instantaneous oscillation together with the achieved steady state amplitudes and B) explicit world view overlay with scientific representation with different degrees of damping runs.

Three advantages of using EJS include 1) Open Source Codes and Creative Commons Attribution Licenses for scaling up of interactively engaging educational practices 2) models made can run on almost any device including Android and iOS and 3) allows for redefining physics educational practices through computer modeling.

Download and unzip for offline use
https://dl.dropboxusercontent.com/u/44365627/lookangEJSworkspace/export/ejss_model_SHMresonance091.zip
Or click this link to run the model
https://dl.dropboxusercontent.com/u/44365627/lookangEJSworkspace/export/ejss_model_SHMresonance091/SHMresonance091_Simulation.xhtml
Source code editable using EJS 5.1 and above:
https://dl.dropboxusercontent.com/u/44365627/lookangEJSworkspace/export/ejss_src_SHMresonance091.zip
2015 resource: http://iwant2study.org/ospsg/index.php/interactive-resources/physics/02-newtonian-mechanics/09-oscillations/88-shm24

I. INTRODUCTION

In the introduction of resonance phenomenon to students, good real life demonstrations, include applying a periodic driving force to swing pendulums of different lengths (Figure 1), illustrate how an oscillatory system will respond with large maximum amplitude when the driving frequency matches its natural frequency. To compliment demonstrations, the Resonance simulation by PhET[1] that shows different masses being oscillated by the same oscillator force could be a useful virtual tool.

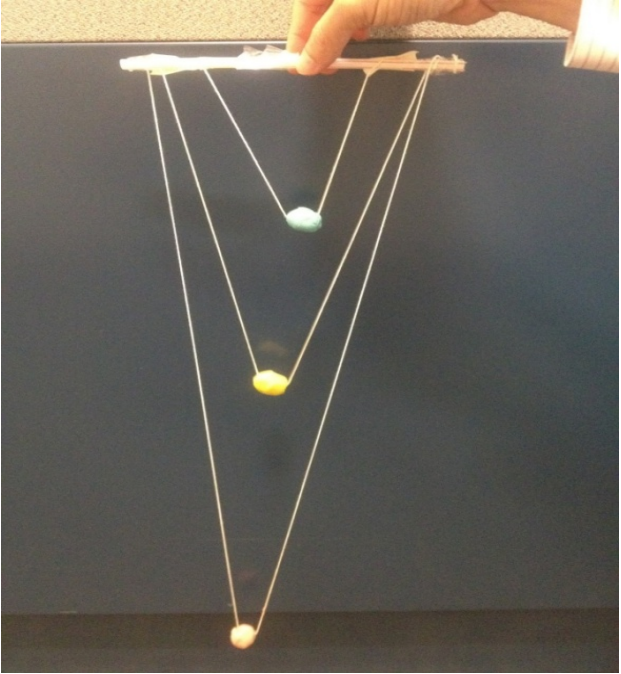

Figure 1. Three equal mass balls attached to a supported structure straw using strings of different lengths to allow students to experience resonance when swinging the masses in and out of the view plane.

We argue that despite such attempts to incorporate hands-on activities with real equipment [2] or computer simulations [1,3-5] in the learning of resonance, most students and teachers may still find the concept of resonance graphs that illustrate the amplitude response of an oscillatory system with varying driving force frequencies difficult [6] to learn and teach. The explicit discussion of (a) the transient-'settling down' and steady state-'time invariant' amplitude and (b) the concept to a single system with one natural frequency to have a response graph (Figure 2) steady state amplitude when subjected to many different driving frequencies is usually missing in most Singapore-Cambridge General Certificate of Education Advanced Level learning activities.

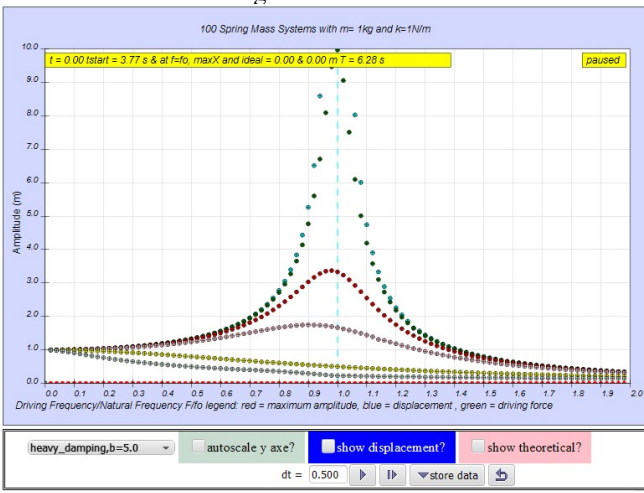

Figure 2. The simulation view of N=100, mass, m = 1kg and spring constant k =1 N/m, of steady state variation of amplitudes with frequency for several damping cases as shown from top to bottom, no damping b=0, very light damping b=0.1, light damping b=0.3, moderate damping, b=0.6, critical damping, b=2.0, and heavy damping, b=5.0 selected arbitrarily for our teaching and learning practices. Run on internet browser:
https://dl.dropboxusercontent.com/u/44365627/lookangEJSworkspace/export/ejss_model_SHMresonance091/SHMresonance091_Simulation.xhtml

Our hypothesis is that by allowing students to visualize the effects of an array of equal spring-masses being acted on by increasing driving frequencies plotted on the traditional "amplitude versus driving frequency" resonance graph, students will gain a greater physical intuition of the resonance concept.

## II. PHYSICS MODEL

Using the Easy Java Simulation (EJS 5.1 [7] authoring toolkit by Francisco Esquembre and has been used by authors [8,9] in Physics Education journal as well), we designed a computer model using an array of 100 identical spring-masses, and subjecting each to an equal driver amplitude but with a small incremental difference in the driver frequency in the x- axis (Figure 2).

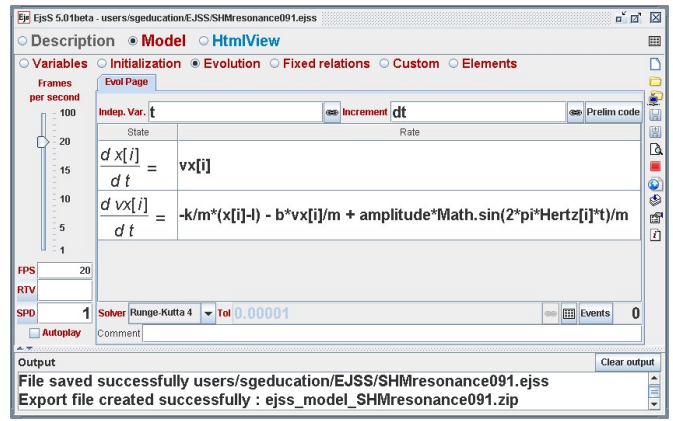

Figure 3. EJS Model Evolution Page with the ordinary differential equation showing how to setup the array of objects with position $x[i]$ and velocity $vx[i]$ and the acceleration relating the spring constant $k$, mass, $m$, damping coefficient $b$ and the driving periodic force of ampitude and $Hertz[i]$ being the driving frequency.

Figure 3 shows how the model is implemented in EJS. An array[$i$] of identical spring-mass system, spring constant $k$ and mass $m$, each having their own positions $x[i]$ and velocities $vx[i]$ is modeled using the differential equations (1) and (2). The evolution of the spring-mass systems with time is achieved by advancing the model by fixed time steps $dt$.

$$\frac{dx[i]}{dt} = vx[i] \quad (1)$$

$$\frac{dvx[i]}{dt} = -\frac{(k)x[i]}{m} - \frac{(b)vx[i]}{m} + \frac{(A)\sin(2\pi(f[i])t)}{m} \quad (2)$$

Equation (1) defines the velocity of each spring-mass system as required by EJS's Model Evolution Page. Equation (2) governs the accelerations acting on each mass that include the forces acted by the spring $F_{spring}(t) = -(k)x[i]$, the effects of damping force $F_{damp}(t) = -(b)vx[i]$ with damping coefficient $b$, and different periodic sinusoidal driving forces $F_{driver}(t) = (A)\sin(2\pi(f[i])t$ with amplitude $A$ and frequency $f[i]$.

A teal color $f_o$ natural frequency vertical dotted line determined using equation (3) of the system is added to aid visualization.

$$f_o = \frac{1}{2\pi}\sqrt{\frac{k}{m}} = \frac{1}{T_o} \quad (3)$$

An option to show the theoretical steady state amplitudes for each driving frequency is represented by the pink dots, is added. These amplitudes, as shown in equation (4), depend only on the driving amplitude $A$, driving frequencies, $f[i]$, and natural frequency $f_o$.

$$x[i] = \frac{(A)\sin\left(2\pi(f[i])t + \tan^{-1}\left(\frac{(2\pi f_o)^2 - (2\pi f[i])^2}{\frac{2(2\pi f[i])(2\pi f_o)b}{2\sqrt{mk}}}\right)\right)}{m\sqrt{\left(2(2\pi f_o)\frac{b}{2\sqrt{mk}}\right)^2 + \frac{1}{\left(\frac{b}{2\sqrt{mk}}\right)^2}((2\pi f_o)^2 - (2\pi f[i])^2)^2 (2\pi f[i])}} \quad (4)$$

To determine an optimum time $t_{startrecording}$ to start recording displacement as maximum amplitudes, we used a minimum of 30 seconds or three times the value of period of free oscillation $T_0$ divide by damping coefficient $b$ as set in equation (5). This serves to allow for accurate tracking of maximum amplitudes for sufficiently long period of time for maximum amplitudes to closely "match" the theoretical solution (Figure 2).

$$t_{startrecording} = \text{Maximum}\left(\frac{3T_0}{b}, 30\right) \quad (5)$$

Note that for the special case where $b = 0$, the systems would have both the non-decaying transient and steady state amplitudes, thus, impossible to reach steady state amplitude regardless of time allowed to run for, therefore we used the theoretical steady state as stored maximum amplitudes values.

### III. HOW TO USE

To use the computer model, students select the desired damping option from the dropdown menu and click on the "play" button. The simulation will compute and display how the 100 equal blue masses are being 'excited' by the periodic force with incrementally increasing frequencies (green arrows, refer to Figure 4, Figure 5, Figure 6 and Figure 7.

In addition, the pink theoretical steady state amplitudes are also plotted. The red maximum amplitudes will be shown a short while after the simulation starts, the teacher can then highlight to the students that maximum amplitudes can be said to be at steady state when most or all of the red points matches the pink points. Click on the "store data" button on the right of the 'reset' button to store this data set. The experiment can be repeated with different damping values by as the masses are re-initialized back to the starting position and the time is reset to $t = 0$ s.

The student can now select the 2$^{nd}$ new damping level say $b = 0.1$ and run the model for a sufficient long time greater than 62.8 s for the maximum amplitudes to be recorded from the masses being excited at different driving frequencies.

The other selectable checkbox on the left allows the $y$ axis to zoom out to 10 to -10, allowing students to visualize the negative part of the displacement. For the typical resonance graph, uncheck this to show the $y$ axes amplitude to show the range from zero to 10 m.

Lastly, the time step $dt = 0.25$ s can be made bigger say to $dt = 1.00$ s for low computing power mobile devices, thus shortening the time and computing resources required to reach a sufficiently long time for maximum amplitudes to be recorded.

### IV. RESULTS OF DIFFERENT DAMPING FORCED RESONANCE GRAPHS

Students can experience the different damping effects (1) Figure 4 for no damping $b=0$, (2) Figure 5 for light damping case $b=0.1$, (3) Figure 6 critical damping $b =2.0$ and (4) Figure 7 heavy damping b = 5.0.

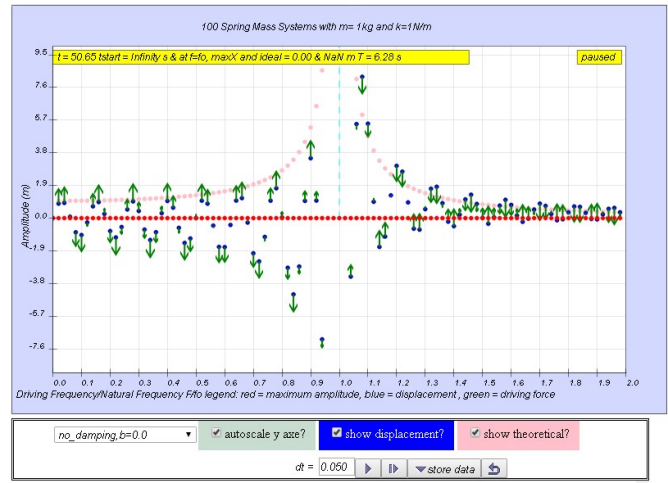

Figure 4. The simulation view of the complete model showing the array of 100 (arbitrarily set) mass (blue color) spring system of mass = 1kg and spring constant = 1 N/m subjected to increasing external driving frequency forces (green color) with the forced resonance graph for no damping case (pink color) storing the theoretical values as maximum amplitudes. Note that autoscale y axes? is checked allowing visualization of the negative displacements.

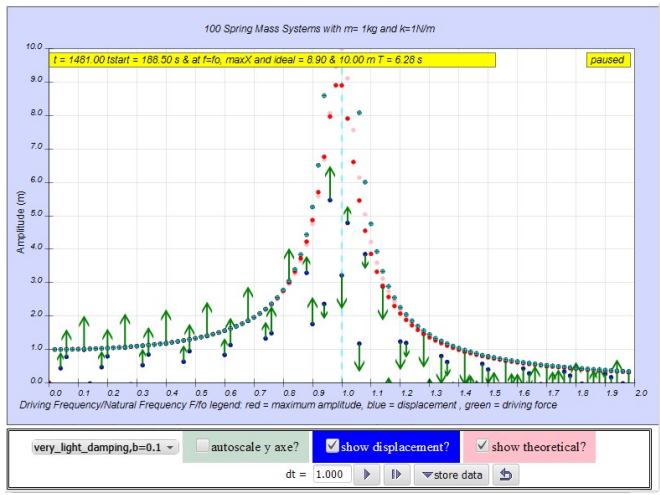

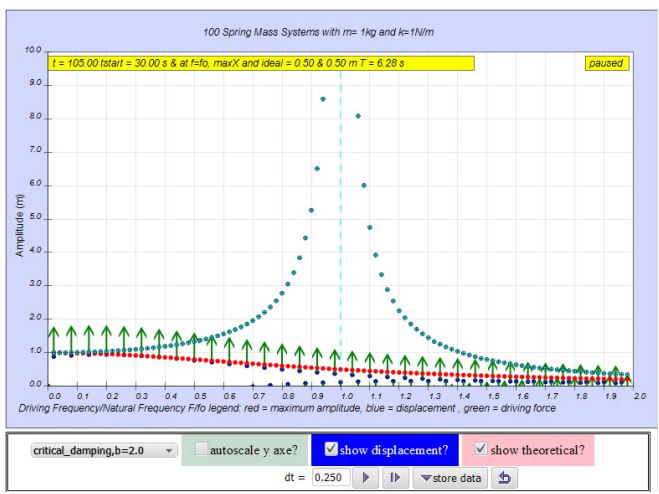

Figure 5. The simulation view of the complete model showing the forced resonance graph for light damping case, b=0.1 (red color showing maximum amplitude calculated after t > 188.5 s) resembling the typically theoretical (pink color) forced resonance line graph when the simulation has run for a sufficiently long time approximating, in this case, t=1481 s for the steady state. Note that the time step dt =1 s reaching steady state faster, especially useful for slower computers and tablets.

Figure 6. The simulation view a complete model no damping case (teal color) and for critical damping case, b =2.0 (red color showing maximum amplitude calculated after t > 30 s) resembling the theoretical (pink color) forced resonance line graph when the simulation has run for a sufficiently long time approximating in this case, t=105.0 s for the steady state.

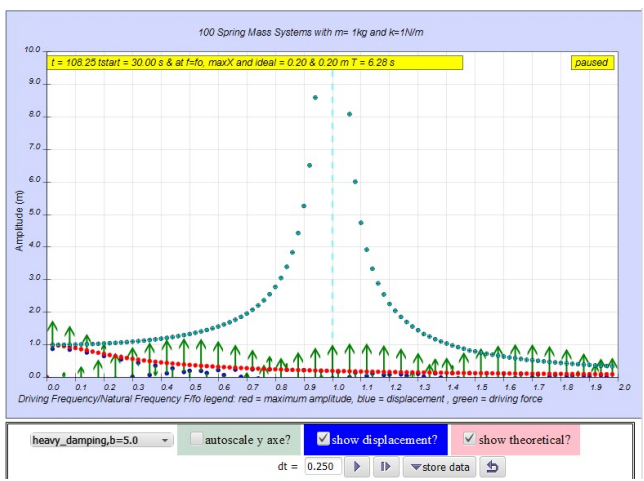

Figure 7. The simulation view of the complete model showing no damping case (teal color) and for heavy damping case b = 5.0 (red color showing maximum amplitude calculated after t > 30 s) resembling theoretical (pink color) forced resonance line graph when the simulation has run for a sufficiently long time approximating in this case, t=108.25 s for the steady state.

Thus, the "store data" feature allows all the different cases to be shown on the same graph for critical viewing and discussions of the steady state amplitudes for different damping.

### V. ADVANTAGES OF THIS COMPUTER MODEL DESIGN

The three most popular simulations or applets on forced resonance phenomena[1,3,4], from Google search, are not designed for the inquiry and derivation of forced resonance amplitude graph with different levels of damping. The Force Oscillation (Resonance)[3,4] simulations by C.K. Ng and Walter Fendt only show the forced resonance amplitude as an approximate[3] or static[4] curve, which still requires students to make implicit links between the real world view and the scientific graphical representation. Even another model[5] by Wolfgang Christian and Loo Kang Wee (Figure 8) shows the different damping by plotting lines which still present novice learners difficulties in interpreting the meaning of these lines.

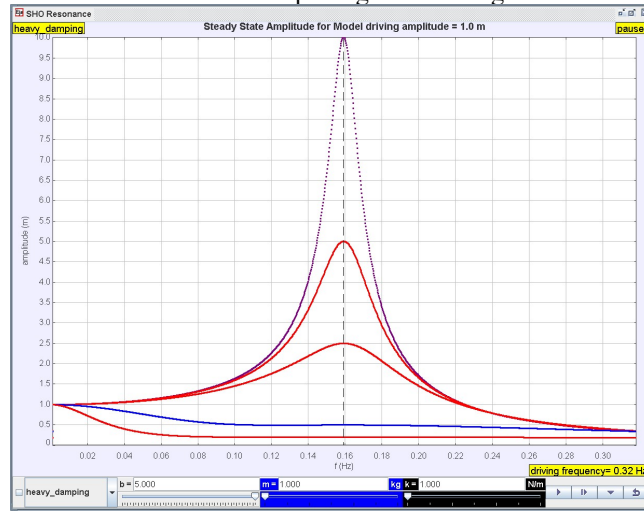

Figure 8. Another EJS Resonance Frequency vs Amplitude Curve Model by Wolfgang Christian (original author) and Loo Kang Wee (sub-author who customize it) but did not show the real world view from which the scientific representations of the forced resonance graph is derived.
Download Simulation with Source:
https://dl.dropboxusercontent.com/u/44365627/lookangEJSworkspace/export/ejs_model_SHOResonancewee.jar

We believe that learning the concept of forced resonance is enriched[10] largely due to the dynamic representation of the physical world view overlaid with the scientific graphical representation [9]made explicit together.

### A. Displaying the transient oscillation together with the maximum & steady state amplitudes

Our model shows the transient instantaneous positions of each mass blue positions being able to 'push' the maximum red amplitudes. This representation aims to allow students to intuitively associate the physical oscillations of the mass with the resulting resonance graph.

### B. Explicit design of computer model with world view coupled with scientific representation with different damping runs

Our model aims to replicate a difficult to setup world view of many spring mass systems being excited by different frequencies driving forces and merge it with the scientific representation[9] with different damping runs for ease of visual comparison. Adding the theoretical steady state (pink) amplitudes as hints for students to decide how long each damping level should be run, we argue further enhanced[10] the dynamic computer model design.

### VI. ADVANTAGES OF USING EASY JAVA SIMULATION

To add to the body of knowledge and literature on the physics model making tools and inform education policy administrators and leaders, we discuss three reasons why using Easy Java Simulation could be an appropriate choice for customization and creating digital library of resources for inquiry based pedagogies that provide interactive engagement.

### A. Open Source Codes and Creative Commons Attribution Licenses

Both the Easy Java Simulation toolkit and the models made with it are open sourced.

Put simply, it means even if the author of EJS, Francisco Esquembre does not continue to develop capabilities for the toolkit, for non-commercial purposes, anyone could potentially continue to develop EJS to benefit users of EJS, where attributing the author Francisco Esquembre and its various contributors is the license associated to its use and redistributions.

Secondly, most if not all of the models created in EJS are have open and freely editable source codes, typically licensed creative commons attribution.

In other words, anyone can remix the existing models created by the various original authors provided the customized models continue to bear the names of the original authors.

### B. Runs on almost any device through internet browser

Easy Java Simulation 5.1, shipped during the Multimedia in Physics Teaching and Learning (MPTL18) conference, Madrid, Spain, September 2013, where we shared our other EJS models [10], allows the creation of models that can be run on almost any device and, computer through the internet browser. In our testing, Chrome, Firefox and Safari are all capable of running these new JavaScript based models, with only the Microsoft Internet Explorer 8 unable to support JavaScript that we have encountered.

### C. Design for the pedagogical practice of modeling

Under the Framework for K-12[11] science education practices, modeling[12] is an essential practice which allows students to articulate, explain and critique their ideas of the natural world and predict the likely behavior of a system. EJS and its growing digital library of physics models support and enrich the practice of developing progressively complex models for computational modeling and thinking.

Even if the computer models are used for interactive engagement[13] as digital resources, which EJS can do because of reasons stated above A and B, it is the redefinition [14] use of technology that transform pedagogical practice through computer modeling that should be scaled up to more schools in preparation of next generation educational systems.

## VII. CONCLUSION

The physics model of forced resonance amplitude versus frequency graphs is discussed in context to EJS, and we recommend allowing academically inclined students to learn physics through the modeling approach [15]as briefly described in II. As a beginning step, perhaps students can be guided through explicitly on what the steps are required to change the EJS model by an EJS experienced teacher before progressing on to more advanced models

For the mainstream students, we recommend using this model as an interactive inquiry learning resource, with III explaining how to use the model.

Our model can produce results in the steady states amplitudes of different damping levels to coincide well with the theoretical presentations and even has the slight leftward shift of the peak as damping increases from no damping to critical damping (Figure 2).

A comparison of our computer model design with three other existing and popular free simulations on the internet and another EJS model suggests this model is probably a world first that shows the world view of many identical spring mass system, each driven by an external periodic sinusoidal force, resulting in a steady state amplitude of different damping forced resonance curves.

Advantages of this computer model include V.A) displaying the instantaneous oscillation together with the steady state amplitudes bridging world view to scientific graphic representation and V.B) explicit design of computer model coupled with scientific representation with different runs of damping levels all shown together for ease of comparison.

Lastly, three advantages of using EJS for furthering physics education include 1) open source codes and creative commons attribution licenses 2) models made using EJS runs of almost any device through internet browser on the operating system of Android, iOS, Windows, MacOS and Linux, and 3) next generation pedagogical practice of modeling is made possible and feasible to all.

## ACKNOWLEDGMENT

We wish to acknowledge the passionate contributions of Francisco Esquembre, Fu–Kwun Hwang, and Wolfgang Christian, Doug Brown, Mario Belloni, Anne Cox, Andrew Duffy, Mike Gallis, Todd Timberlake, Taha Mzoughi and many more in the Open Source Physics community for their ideas and insights in the co–creation of interactive simulation and curriculum materials.

| | AUTHORs |
|---|---|
| 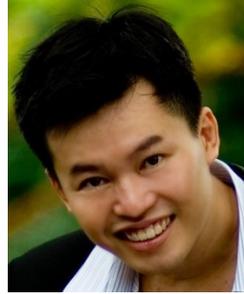 | Loo Kang WEE Lawrence is currently educational technology specialist II at the Ministry of Education, Singapore. His open source physics contribution garner awards including Public Service PS21 -Distinguished Star Service Award 2014 & Best Ideator 2012, Ministry of Education-Best Innovator Award 2013 & Excellence Service Award 2012. |
| 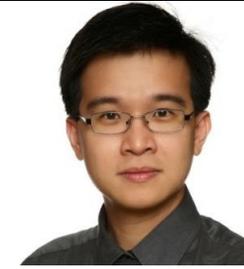 | Tat Leong LEE is currently an educational technology officer. He has been using Open Source Physics (OSP) tools as early as 2006 (Tracker and Easy Java Simulations). |
| 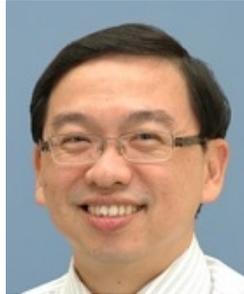 | Dr Charles CHEW is currently a Principal Master Teacher (Physics) with the Academy of Singapore Teachers. He has a wide range of teaching experiences and mentors many teachers in Singapore. He is an EXCO member of the Educational Research Association of Singapore (ERAS) and is active in research to strengthen theory-practice nexus for effective teaching and meaningful learning. |
| 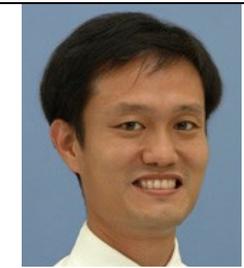 | Dr Darren Wong is currently a senior curriculum specialist at the Ministry of Education, Singapore. His current research interests are in the areas of physics education research, inquiry-based learning in science and teacher professional development. |
| 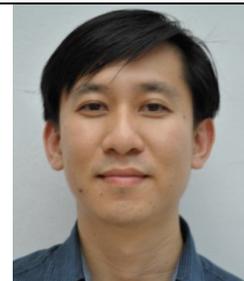 | Samuel TAN is an educational technology specialist at the Ministry of Education, Singapore. He designs ICT-enabled learning resources for Science and develops ICT pedagogical frameworks for implementation at schools. |